\documentclass[showpacs,aps,prl,twocolumn,superscriptaddress,footinbib]{revtex4}
\usepackage{graphicx} 
\usepackage{dcolumn}
\usepackage{bm}

\begin{document}

\title{Nonlinear Photoluminescence Excitation Spectroscopy of Carbon Nanotubes: \\Exploring the Upper Density Limit of One-Dimensional Excitons}

\author{Yoichi Murakami}
\affiliation{Department of Electrical and Computer Engineering, Rice University, Houston, Texas 77005, USA}
\affiliation{Department of Chemical System Engineering, University of Tokyo, 
Bunkyo-ku, Tokyo 113-8656, Japan}

\author{Junichiro Kono}
\email[]{kono@rice.edu}
\thanks{corresponding author.}
\affiliation{Department of Electrical and Computer Engineering, Rice University, Houston, Texas 77005, USA}


\begin{abstract}
We have studied emission properties of high-density excitons in single-walled carbon nanotubes through nonlinear photoluminescence excitation spectroscopy.  As the excitation intensity was increased, all emission peaks arising from different chiralities showed clear saturation in intensity.  Each peak exhibited a saturation value that was independent of the excitation wavelength, indicating that there is an upper limit on the exciton density for each nanotube species.  We developed a theoretical model based on exciton diffusion and exciton-exciton annihilation that successfully reproduced the saturation behavior, allowing us to estimate exciton densities.  These estimated densities were found to be still substantially smaller than the expected Mott density even in the saturation regime, in contrast to conventional semiconductor quantum wires.

\end{abstract}


\maketitle

Linear and nonlinear optical processes in one-dimensional (1-D) semiconductors are affected by strong Coulomb interactions and have been the subject of many theoretical~\cite{OgawaTakagahara91both,RossiMolinari96PRL,TassonePiermarocchiBoth,DasSarmaWangBoth} and experimental~\cite{KaponetAl89PRL,WegscheideretAl93PRL,AmbigapathyetAl97PRL,RubioetAl01SSC,AkiyamaetAl03PRB,GuilletetAl03PRB,HayamizuetAl07PRL} studies.
Early studies of lasing in semiconductor quantum wires (QWRs)~\cite{KaponetAl89PRL,WegscheideretAl93PRL} stimulated much interest in the fundamental properties of {\em high-density} 1-D excitons, or correlated electron-hole ($e$-$h$) pairs, especially in their stability against biexciton formation, band-gap renormalization (BGR), and dissociation.  Although it has been established that excitons in QWRs are stable up to very high densities (10$^5$-10$^6$~cm$^{-1}$), unanswered questions still remain.  As the density increases, an insulating exciton gas is expected to become unstable and eventually transform into a metallic $e$-$h$ plasma at the Mott density, where the inter-exciton distance approaches the exciton size.  At what density gain should appear and whether a clear Mott transition exists in 1-D systems are open questions both theoretically~\cite{RossiMolinari96PRL,TassonePiermarocchiBoth,DasSarmaWangBoth} and experimentally~\cite{AmbigapathyetAl97PRL,RubioetAl01SSC,AkiyamaetAl03PRB,GuilletetAl03PRB,HayamizuetAl07PRL}.

Single-walled carbon nanotubes (SWNTs)~\cite{Iijima-Nature-1993} have recently emerged as novel 1-D solids with very strong quantum confinement.
The exciton binding energies reported for semiconducting SWNTs are very large ($\sim$400~meV~\cite{Wang-Science-2005, Maultzsch-PRB-2005}) compared to typical GaAs QWRs ($\sim$20~meV~\cite{WegscheideretAl93PRL,RossiMolinari96PRL}).  Despite much recent progress in understanding their basic optical properties, so far only a limited number of studies have been performed under the condition of high carrier/exciton densities~\cite{Wang-PRB-2004,Ostojic-PRL-2005,Ma-PRL-2005}.
In particular, there have been no reports quantifying exciton densities in relation to the Mott density.

Here we describe detailed characteristics of photoluminescence (PL) and photoluminescence excitation (PLE) spectra of SWNTs at high exciton densities.  Although PL intensity rapidly saturated with increasing pump fluence, PL spectra were stable, indicating that there is no BGR, screening, or dissociation into an $e$-$h$ plasma.  However, we observed {\em significant broadening and eventual complete flattening of PLE spectra} at high laser intensities.  We show that this originates from strong exciton-exciton annihilation (EEA) that provides an upper limit for the exciton population in the lowest-energy ($E_{11}$) state.  Our model, taking into account the diffusion length of excitons in the EEA process, allowed us to estimate exciton densities from experimental PL saturation curves.  The estimated densities for the highly saturated regime were on the order of 1~$\times$~10$^5$~cm$^{-1}$, which remains more than one order of magnitude smaller than the expected Mott density and explains the observed stability of PL spectra even in the saturation regime.

\begin{figure}
\includegraphics[scale=0.48]{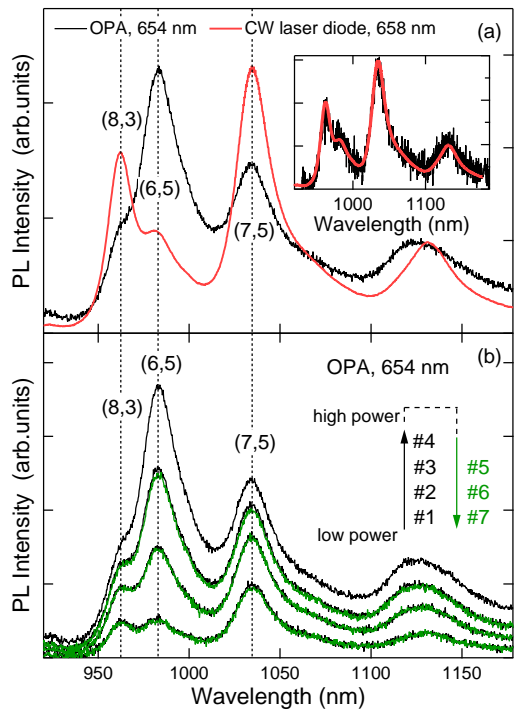}
\caption{(color online). Pump-intensity-dependent PL spectra. (a)~Black --- spectrum obtained with OPA (654~nm, 29~nJ).  Red curve --- spectrum obtained with a CW laser diode (658~nm, 100~$\mu$W).  Inset shows that the two spectra coincide when the OPA pulse energy is very low (300~pJ).  (b)~Change of PL spectra versus OPA pulse energy between 1~nJ and 30~nJ (in the order of \#1 to \#7).  Curve \#4 corresponds to the highest fluence ($\sim$1.3~$\times$~10$^{14}$~photons/cm$^2$).
}
\label{PL}
\end{figure}

The sample we studied was a centrifuged supernatant of CoMoCAT SWNTs dispersed by 1~wt\% sodium cholate in D$_2$O.
The solution was held in a 1-mm-thick quartz cuvette.  The optical density in the $E_{22}$ region was below 0.2, which helped avoid inhomogeneous excitation and re-absorption of PL.  Our excitation source was a 1~kHz, $\sim$250~fs optical parametric amplifier (OPA), tunable in the visible and near-infrared, pumped by a chirped pulse amplifier (Clark-MXR, CPA2010).  
The OPA beam was focused onto the sample with a spot size of 300-400~$\mu$m.
The PL from the sample was recorded with a liquid-nitrogen-cooled InGaAs array detector.

Figure 1(a) compares two PL spectra.  The black curve represents excitation by the OPA with a wavelength of 654~nm (or 1.90~eV) and a pulse energy of 29~nJ, while the red curve represents excitation by a weak (100~$\mu$W) CW laser with a wavelength of 658~nm (or 1.88~eV).  It is seen that the relative intensities of different PL peaks are drastically different between the two curves.  The inset confirms that the two spectra actually coincide very accurately when the OPA pulse energy is kept very low (300~pJ).  Figure~\ref{PL}(b) shows PL spectra for pulse energies of 1~nJ (curves \#1 and \#7), 4~nJ (\#2 and \#6), 10~nJ (\#3 and \#5), and 30~nJ (\#4).  The (7,5) peak is dominant at low fluences while the (6,5) peak becomes dominant at high fluences.  It is important to note that the different curves were taken in the order of \#1 to \#7, demonstrating that the observed changes are reproducible and are not caused by any laser-induced permanent change in the sample.  We also reproduced a similar power dependence in a dried film, showing that the power dependent changes are not an artifact caused by the fluidic nature of the sample.  Additionally, note that the PL intensities tend to saturate at high laser fluences, while their peak positions do not change at all and linewidths increase only slightly (by $\sim$15\% from curve \#1 to \#4). 

\begin{figure}
\includegraphics[scale=0.54]{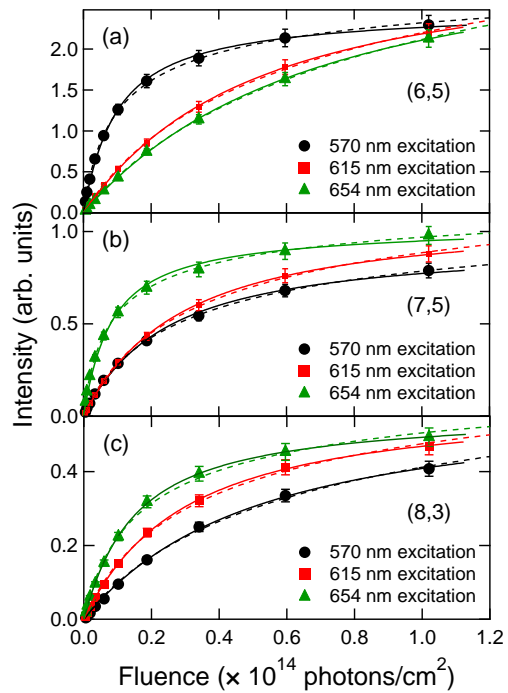}
\caption{(color online). Integrated PL intensity versus pump fluence for (6,5), (7,5), and (8,3) SWNTs.  Pump wavelengths were 570~nm (circles), 615~nm (squares), and 654~nm (triangles).  The error bars account for $\pm$5\%.  The solid and dashed curves are fitting by Eq.~(1) and Eq.~(2), respectively.}
\label{saturation}
\end{figure}

Figures \ref{saturation}(a)-\ref{saturation}(c) plot integrated PL intensities for (6,5), (7,5), and (8,3) tubes, respectively, versus pump fluence up to $\sim$1.02~$\times$~10$^{14}$~photons/cm$^2$ with pump wavelengths of 570, 615, and 654~nm, showing clear saturation behaviors at high pump fluences.  The $E_{22}$-resonant wavelengths of these tube types are 570, 647, and 673~nm, respectively.  The integrated PL intensities were obtained through spectral decomposition analysis
by assuming 50\% Lorentzian + 50\% Gaussian, while keeping all the peak positions fixed.
Figure~\ref{saturation} indicates that saturation starts at a lower (higher) fluence when the sample is resonantly (non-resonantly) excited.  The unexpectedly fast saturation of the (7,5) peak with 570~nm excitation (which is non-resonant) is likely due to its proximity to the phonon sideband at 585~nm~\cite{Miyauchi-PRB-2006}.  
The solid and dashed curves are theoretical fits to the experimental curves based on our model to be discussed later.

\begin{figure}
\includegraphics[scale=0.55]{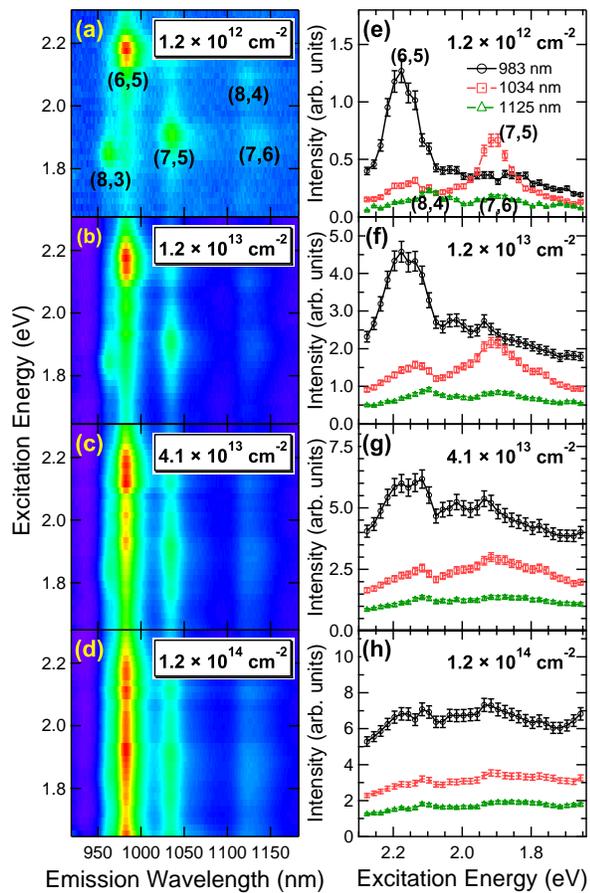}
\caption{(color online).  Evolution of PLE data with increasing pump pulse fluence: (a) 1.2 $\times$ 10$^{12}$, (b) 1.2 $\times$ 10$^{13}$, (c) 4.1 $\times$ 10$^{13}$, and (d) 1.2 $\times$ 10$^{14}$ photons/cm$^2$.
(e-h): PLE spectra corresponding to (a)-(d) at emission wavelengths of 983~nm (circles), 1034~nm (squares), and 1125~nm (triangles).}
\label{PLE}
\end{figure}

Figures \ref{PLE}(a)-\ref{PLE}(d) show PLE maps taken with various pump fluences.  The step size for the pump photon energy was 20~meV.  The data taken with the lowest fluence (1.2~$\times$~10$^{12}$ photons/cm$^2$) [Fig.~\ref{PLE}(a)] is essentially the same as that taken with low-intensity CW light.  However, as the fluence is increased [Figs.~\ref{PLE}(b)-\ref{PLE}(d)], the $E_{22}$ excitation peaks gradually broaden and eventually become completely flat at the highest fluence (1.2~$\times$~10$^{14}$~photons/cm$^2$) --- i.e., {\em PL intensities become independent of the excitation wavelength}.  The corresponding PLE spectra are shown in Figs.~\ref{PLE}(e)-\ref{PLE}(h) for three PL wavelengths at 983, 1034, and 1125~nm.  

\begin{figure}
\includegraphics[scale=0.5]{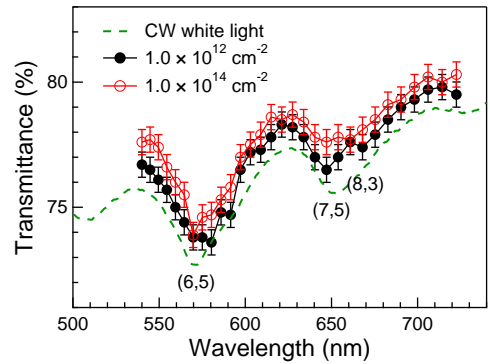}
\caption{(color online). Transmission spectra measured with high-fluence OPA pulses (filled circles: 1.0 $\times$ 10$^{12}~$photons/cm$^{2}$, open circles: 1.0 $\times$ 10$^{14}$~photons/cm$^{2}$) compared with a transmission spectrum taken with a weak CW white-light beam (dashed line).}
\label{abs}
\end{figure}

To provide further insight into the nature of the observed broadening and flattening of PLE spectra, we also performed absorption measurements in the $E_{22}$ region using OPA pulses.  Figure~\ref{abs} compares two transmission spectra measured with the OPA at fluences of 1.0~$\times$~10$^{12}$~photons/cm$^2$ (linear regime) and 1.0~$\times$~10$^{14}$~photons/cm$^2$ (saturation regime), respectively, and a transmission spectrum taken with weak CW whitelight.
It is clear from the figure that the $E_{22}$ absorption peaks do not exhibit any broadening and shifts even in the saturation regime, strongly suggesting that these intensities are not high enough to cause light-induced nonlinear effects such as state filling, carrier-density-dependent dephasing, and non-perturbative light-matter coupling (or ``dressing'') on states in the $E_{22}$ range.

We interpret these observations as results of very efficient EEA~\cite{Roberti-JCP-1998,Ma-PRL-2005}.  We assume that the formation of $E_{11}$ excitons occurs in a very short time scale (e.g., $\sim$40~fs~\cite{Manzoni-PRL-2005}), shorter than our pulse width ($\sim$250~fs).  Thus, excitons quickly accumulate in the $E_{11}$ state during and right after photo-creation of $e$-$h$ pairs.  However, the number of excitons that can be {\em accommodated} in the $E_{11}$ state is limited by EEA.  As the exciton density, $n_{\rm x}$, approaches its maximum value, EEA begins to prevent a further increase by {\em efficiently removing excitons non-radiatively}, which explains the PL saturation behavior.  Since EEA serves as a bottleneck for the exciton density, the PL intensity becomes insensitive to whether the excitons were created resonantly or non-resonantly around the $E_{22}$ level and independent of the pump wavelength.  Namely, at very high pump fluences, the PL intensity is determined not by how efficiently excitons are created but by {\em how many $E_{11}$ excitons can be accommodated within a particular type of SWNT} as well as by the relative abundance of that type of SWNT in the sample.

To determine exciton densities $n_{\rm x}$ from the PL saturation curves, we have developed a theoretical model~\cite{Murakami-in-preparation}.
Existing EEA models~\cite{Roberti-JCP-1998, Ma-PRL-2005} assume the annihilaiton rate to be $\propto n_{\rm x}^2$,
a probability of finding two excitons at the same position in space.
However, its validity becomes questionable when one deals with high density excitons in 1D where the dynamics and/or size of the excitons are expected to influence the annihilation rate.
We introduce a dimensionless exciton population ($0 \leq \zeta < 1$) defined as $\zeta \equiv N\/l_{\rm x} / L_{\rm{NT}}$, where $N$ is the number of excitons in a SWNT, $L_{\rm{NT}}$ is the length of the SWNT, and $l_{\rm x}$ is the characteristic length scale each exciton ``occupies'' in the SWNT.  In a static limit, $l_{\rm x}$ should simply be the exciton size.  However, recent experiments have indicated the importance of exciton {\em diffusion}~\cite{Sheng-PRB-2005, Russo-PRB-2006,Cognet-Science-2007}.  In particular, Cognet {\em et al}.~\cite{Cognet-Science-2007}, via micro-PL studies on single micelle-suspended SWNTs of various chiralities, found that excitons diffusively traverse a substantial distance ($\sim$90~nm) within their lifetime.  Therefore, we assume that $l_{\rm x}$ is determined by the diffusion length and any two excitons formed within $l_{\rm x}$ undergo EEA.  We also assume that an exciton promoted to a higher energy level as a result of EEA returns to the $E_{11}$ level with 100\% probability due to ultrafast and efficient $E_{22}$-to-$E_{11}$ relaxation; i.e., two excitons become one exciton through EEA.

Two limiting cases are considered, which we refer to as the ``instantaneous'' and ``steady-state'' limits. In the instantaneous limit, the initial creation of excitons is completed before EEA follows, while the steady-state limit corresponds to CW excitation.  The actual experimental situation is considered to be closer to the former.  Assuming $L_{\rm{NT}} \gg l_{\rm x}$, the PL saturation equation is given~\cite{Murakami-in-preparation} for the instantaneous limit
\begin{equation}
\psi = \frac{1}{e}\!\left\{ \mathrm{Ei}\!\left(\frac{\zeta}{1-\zeta}\right) - \mathrm{Ei}\!\left(1\right) \right\},
\label{instantaneous-eqn}
\end{equation}
and for the steady-state limit
\begin{equation}
\psi = \frac{\zeta}{1-\zeta} \, \mathrm{exp}\left(\frac{\zeta}{1-\zeta} \right) \;\;
\left[\zeta \equiv \frac{I_{\mathrm{PL}}}{c_1}, \; \psi \equiv \frac{I_{\mathrm{pump}}}{c_2} \right].
\label{steady-state-eqn}
\end{equation}
These are implicit equations relating the PL intensity ($I_{\rm{PL}}$) and pump intensity ($I_{\rm{pump}}$) in terms of their respective dimensionless parameters, $\zeta$ and $\psi$.  They contain no fitting parameters other than the two linear scaling factors $c_1$ and $c_2$ and simply become a linear relationship ($\psi = \zeta$) in the low density limit ($\zeta \rightarrow 0$).  The solid and dashed curves in Fig.~\ref{saturation} are fits to the data by Eqs.~(1) and (2), respectively, showing good agreement.  The values of $\zeta$ were obtained through this analysis for respective data points shown in Fig.~\ref{saturation}.  Finally, the exciton density can be obtained through $n_{\rm{x}}$ = $\zeta$/$l_{x}$.

Using $l_{x}$ = 45~nm (one half of the excursion range in~\cite{Cognet-Science-2007}), we estimated $n_{\rm x}$ for (6,5) to be 1.7 $\times$ 10$^5$, 1.3 $\times$ 10$^5$, and 1.1 $\times$ 10$^5$~cm$^{-1}$ for excitation with 570, 615, and 654~nm, respectively, at a fluence of 1.02 $\times$ 10$^{14}$ photons/cm$^2$ in the instantaneous limit.  These are much smaller than the expected Mott density $n_{\rm x}^*$ ($\sim$7~$\times$~10$^6$~cm$^{-1}$, assuming exciton size $\sim$~1.5~nm~\cite{Perebeinos-PRL-2004}), which explains the stability of PL (Fig.~\ref{PL}) in the saturated regime.  However, the densities of {\em as-created} $e$-$h$ pairs is estimated to be 1-2~$\times$~10$^6$~cm$^{-1}$ in the cases of resonant excitation, which is similar to $n_{\rm x}^*$.   These observations appear qualitatively different from GaAs QWRs where PL saturation is not obvious until the formation of biexcitons and an $e$-$h$ plasma~\cite{HayamizuetAl07PRL}.  Highly efficient {\em and} rapid EEA in SWNTs, which is consistent with the reported absence of biexciton signatures~\cite{Matsuda-PRB-2008}, is probably the direct reason for such characteristic differences.

In summary, we have studied PL and PLE spectra of SWNTs at high exciton densities.  Complete flattening of PLE spectra and clear PL saturation were observed, indicating the existence of an upper density limit.  We developed a model that reproduced the observed saturation behavior and allowed us to estimate exciton densities, which remained more than one-order-of-magnitude lower than the expected Mott density and explained the stability of PL spectra in the saturated regime.  

\begin{acknowledgments}
The authors thank the Robert A.~Welch Foundation (Grant No.~C-1509) and NSF (Grant No.~DMR-0325474) for support and H.~Akiyama and K.~Matsuda for valuable discussions.  One of us (Y.M.) thanks T.~Okubo and S.~Maruyama for their support for the fulfillment of his JSPS program \#18-09883 at Rice University.
\end{acknowledgments}


\end{document}